\documentclass[twocolumn,prd,nofootinbib,aps,floats,floatfix,amsmath,amssymb,secnumarabic]{revtex4} %

\usepackage{graphicx}
\usepackage[usenames,dvipsnames]{color}
\usepackage{amsmath,amssymb}
\usepackage{slashed}
\usepackage[colorlinks,citecolor=blue]{hyperref}

\def\sfrac#1#2{{\textstyle{#1\over #2}}}
\newcommand{\be}{\begin{equation}}
\newcommand{\ee}{\end{equation}}
\newcommand{\ba}{\begin{array}}
\newcommand{\ea}{\end{array}}
\newcommand{\bea}{\begin{eqnarray}}
\newcommand{\eea}{\end{eqnarray}}

\newcommand{\lrasd}{%
  \mathrel{\vbox{\offinterlineskip\ialign{%
    \hfil##\hfil\cr
    $\scriptscriptstyle\leftrightarrow$\cr
    $\slashed{\partial}$\cr
}}}}

\begin{document}
\title{Inert Doublet Dark Matter with Strong Electroweak Phase Transition}
\author{Debasish Borah}
\email{debasish@phy.iitb.ac.in}
\affiliation{Department of Physics, Indian Institute of Technology Bombay, 
Mumbai - 400076, India}
\author{James M.\ Cline}
\email{jcline@physics.mcgill.ca}
\affiliation{Department of Physics, McGill University,
3600 Rue University, Montr\'eal, Qu\'ebec, Canada H3A 2T8}
\begin{abstract}

We reconsider the strength of the electroweak phase transition  (EWPT)
in the inert doublet dark matter model, using a quantitatively
accurate form for the one-loop finite temperature effective potential,
taking into account relevant particle physics and dark matter
constraints, focusing on a standard model Higgs mass near 126 GeV,
and doing a full scan of the space of otherwise unconstrained
couplings.  We find that there is a significant (although fine-tuned)
space of  parameters
for achieving an EWPT sufficiently
strong for baryogenesis while satisfying the XENON100 constraints
from direct detection and not exceeding the correct 
 thermal relic density.
We predict that the dark matter mass should be in the range $60-67$ GeV,
and we discuss possible LHC signatures of the charged and CP-odd
Higgs bosons, including a $\sim 10$\% reduction of the $h\to \gamma\gamma$
branching ratio.

\end{abstract}
\pacs{}
\maketitle

\section{Introduction}

Models of scalar dark matter (DM) $H$ can have interesting connections
to Higgs boson ($h$) physics because of the dimension 4 operator
$|h|^2|H|^2$.  One obvious consequence is the possibility of the
invisible decay channel $h\to HH$ if the dark matter is
sufficiently light.  Another is that such a
coupling can allow the electroweak phase transition (EWPT) to become
first order, and potentially strong enough to be interesting for
baryogenesis \cite{Anderson:1991zb}.  There has been considerable
interest in the interplay between dark matter and the
electroweak phase transition in recent years 
\cite{Dimopoulos:1990ai}-\cite{Gonderinger:2012rd}.

The Inert Doublet Model (IDM) is a widely studied setting for scalar
dark matter \cite{Ma:2006km,Barbieri:2006dq,Majumdar:2006nt,LopezHonorez:2006gr}
that can have rich phenomenological consequences 
\cite{Cao:2007rm}-\cite{Arhrib:2012ia}.  Recently its capacity
for giving a strong EWPT was considered by ref.\ 
\cite{Chowdhury:2011ga}.  That work found a rather large allowed
region of parameter space where the EWPT could be strong and other
constraints satisfied, including the correct thermal relic DM density.
However, it employed a simplified version of the finite-temperature
effective potential, keeping only terms up to $O(m/T)^3$ in the
high-temperature expansion.   On the other hand, a very quantitative
treatment of the effective potential for two-Higgs doublet models was
recently undertaken in ref.\ \cite{Cline:2011mm}.  Our purpose in this paper is
to reexamine the strength of the EWPT in the IDM using this more
accurate potential.  Moreover we search the full parameter space of 
the model using Monte Carlo methods, rather than a restricted subspace
using a grid search as was done in \cite{Chowdhury:2011ga}.  We also focus on values
of  the standard model-like Higgs boson mass near 126 GeV, the value 
favored by recent LHC data \cite{ATLAS:2012ae}-\cite{ATLAS:2012ad}.

In this way we are able to extend the results of \cite{Chowdhury:2011ga}, confirming that
there exists a significant (though finely tuned) region of
 parameter space in the IDM where the 
strength of the EWPT is sufficiently enhanced for electroweak
baryogenesis while satisfying other necessary constraints.  The paper
is structured as follows.  
We review the definition of
the model and collider mass constraints in section \ref{model}, our
methodology for defining the effective potential and scanning the
parameter space in section \ref{method},  and we present the results
of the Monte Carlo search in section \ref{results}.  Prospects for
testing the model at colliders are discussed in section \ref{lhc},
and we give conclusions in section \ref{conclusions}.

\section{The Inert Doublet Model}
\label{model}

The Inert Doublet Model is the extension of the standard model (SM) 
by an additional Higgs doublet $S$ with the discrete $Z_2$ symmetry
$S\to -S$, which  naturally leads to a stable dark
matter candidate in in one of the components of $S$ 
\cite{Barbieri:2006dq,Cirelli:2005uq,LopezHonorez:2006gr}. Since an
unbroken $Z_2$ symmetry forbids Yukawa couplings involving $S$, the second
doublet interacts with the SM fields only through
its couplings to the SM Higgs doublet and the gauge bosons. The
scalar potential of the IDM is given by
\begin{eqnarray}
V &=& \frac{\lambda}{4}\left(H^{\dagger i}H_i 
-\frac{v^2}{2}\right)^2 + m^2_1 (S^{\dagger i}S_i)\nonumber\\
 &+& \lambda_1 (H^{\dagger i}H_i)(S^{\dagger j}S_j)+\lambda_2 
(H^{\dagger i}H_j)(S^{\dagger j}S_i) \nonumber \\
&+&[\lambda_3 H^{\dagger i}H^{\dagger j} S_iS_j +\text{h.c.}] + 
\lambda_S (S^{\dagger i}S_i)^2
\end{eqnarray}

We assume that $S$ does not 
acquire a vacuum expection value (VEV),
so as to keep the  $Z_2$ symmetry unbroken.  
The tree-level scalar mass eigenvalues are 
\begin{eqnarray}
m^2_h &=& \sfrac12\lambda v^2 \nonumber\\
 m^2_H &=& m^2_1 +
\sfrac12(\lambda_1+\lambda_2+2\lambda_3)v^2\nonumber\\
m^2_A &=& m^2_1 +
\sfrac12(\lambda_1+\lambda_2-2\lambda_3)v^2\nonumber\\
 m^2_\pm &=& m^2_1 + \sfrac12\lambda_1 v^2
\label{treemass}
\end{eqnarray}
where $m_h$ is the SM-like Higgs mass, $m_H$ $(m_A)$ is the mass of
CP-even (odd) component of the inert doublet, and $m_\pm$ is the 
mass of the charged Higgs.  Without loss of generality, we can take
$\lambda_3 <0$ so that $m_H < m_A$ and therefore $H$ is the dark 
matter particle.  The case $\lambda_3>0$ just corresponds to renaming
$H\leftrightarrow A$. We further restrict 
$\lambda_2+2\lambda_3 <0$ so that $m_H < m_\pm$ to avoid the charged
state being dark matter.

\subsection{Collider Mass Bounds}
\label{cmb}
Precision measurement of the $Z$ boson decay width at LEP I forbids the $Z$
boson decay channel $Z \rightarrow H A$, which requires that $m_H + m_A >
m_Z$. In addition, LEP II constraints roughly rule out the triangular
region \cite{Lundstrom:2008ai}
\[
	m_{H} < 80\ {\rm\ GeV},\quad m_{A} < 100{\rm\ GeV},\quad
	m_{A} - m_{H} > 8{\rm\ GeV}
\]
We take the lower bound on the charged scalar mass $m_\pm > 90
\rm{\ GeV}$ \cite{Pierce:2007ut}. Following the recent LHC exclusion 
of SM-like Higgs masses in the region $127-600 \; \text{GeV}$ 
\cite{ATLAS:2012ae}-\cite{ATLAS:2012ad}, we restrict $m_h$ to the window
$115-130 \; \text{GeV}$, with special attention to the currently 
favored value $m_h\cong 126$ GeV.

\section{Methodology}
\label{method}

In this work we employ the Landau-gauge one-loop finite-temperature
effective potential similar to that described in ref.\
\cite{Cline:2011mm}, which considered the most general two-Higgs
doublet potential.  It has the zero-temperature one-loop corrections
and counterterms to insure that tree-level mass and VEV relations are
preserved, and includes contributions from the scalars, vectors,
Goldstone bosons, and the top quark.  It further implements
resummation of thermal masses.  

As in \cite{Cline:2011mm}, we search the full parameter space of the
model using a Markov chain Monte Carlo (MCMC).  Models are chosen in such a
way as to  favor those with large values of  $(v_c/T_c)$, the ratio of
the Higgs VEV to the critical temperature,  which is the figure of
merit for a strong electroweak phase transition, for the purposes of
electroweak  baryogenesis (see ref.\ \cite{Cline:2006ts} for a review).   In addition, we favor models with small
values of $\lambda_{DM}$, the effective coupling of the DM to the
Higgs boson,
\be
	\lambda_{DM} = (\lambda_1+\lambda_2+2\lambda_3)
\label{ldmeq}
\ee 
since this is required by the XENON100 direct
detection constraint (see below).  Therefore we bias the MCMC using
the combination $a \equiv (v_c/T_c)/\lambda_{DM}$, which is designed
to produce chains of models such that the probability distribution 
$dP/da$, treating the chain as a statistical ensemble, goes like $a$.  We took the random
step size for each of the free parameters of the potential to be
$\sim 10$\% of their starting values, determined by a seed model that
satisfied the constraints enumerated next.  

We implemented the relevant phenomenological and consistency
constraints as in \cite{Cline:2011mm}, namely precision electroweak
observables (EWPO), collider mass bounds as described in section
\ref{cmb}, vacuum stability, and the absence of Landau poles below 2
TeV (an arbitrary cutoff, but sufficient for  considering the model to
be a valid effective theory up to reasonably high energies). 

For the present study, we add to the above criteria the  requirement of the
correct thermal relic density of dark matter.
The relic abundance of a dark matter particle $H$ is given by  
\cite{Jungman:1995df,Jarosik:2010iu}
\begin{equation}
\Omega_{H} h^2 \approx \frac{3 \times 10^{-27} \; \text{cm}^3 \text{s}^{-1}}{\langle \sigma v \rangle}
	= 0.1123 \pm 0.0035
\label{wmap}
\end{equation}
Depending on the DM mass $m_{DM}$, different annihilation
channels contribute to the thermally averaged annihilation cross section.  Here we
consider $m_{DM} < m_W$ so that the only relevant annihilation
channels are those which are mediated by the SM-like Higgs boson
into final state $f\bar f$ pairs (excluding the top quark). Specifically, we
consider the dark matter mass window $45-80 \; \text{GeV}$ (to be justified by the
results below), and the annihilation
cross section
\be
	\langle \sigma v\rangle = \sum_f {3\lambda^2_{DM} m^2_f \over 
	4\pi\left( (4 m^2_{DM}-m^2_h)^2 + \Gamma_h^2 m_h^2\right)}
\label{sigma_relic}
\ee
where $m_{DM} = m_H$, $\Gamma_h \cong 0.003$ GeV is the decay 
width of the 
Higgs boson (at $m_h \cong 126$ GeV), $\lambda_{DM}$ is given 
by (\ref{ldmeq}), and the sum is over all
kinematically accessible SM fermions, thus dominated by $b\bar b$ pairs in the
final state.

In addition to the WMAP constraints on relic density, there is also a strict 
limit on the spin-independent dark matter-nucleon cross section coming from
direct detection experiments, notably XENON100 
\cite{Aprile:2010um}-\cite{Aprile:2012nq}. The
relevant cross section in the present model is given
by \cite{Barbieri:2006dq}
\begin{equation}
 \sigma_{SI} = \frac{\lambda^2_{DM}f^2}{4\pi}\frac{\mu^2 m^2_n}{m^4_h m^2_{DM}}
\label{sigma_dd}
\end{equation}
where $\mu = m_n m_{DM}/(m_n+m_{DM})$ is the DM-nucleon reduced mass.
The Higgs-nucleon coupling $f$ is subject to hadronic uncertainties
that have been discussed in 
refs.\ \cite{Ellis:2008hf,Bottino:2008mf}, and more recently in
\cite{Mambrini:2011ik}; in particular the quark matrix element 
$\sigma_{\pi N}$ upon which $f$ depends is poorly determined.
Many authors take $f\sim 0.35$; for example 
DarkSUSY \cite{Gondolo:2004sc} uses $f=0.38$, while ref.\
\cite{Ellis:2000ds} finds $f=0.35$ and a recent estimate based on
lattice gauge theory  \cite{Giedt:2009mr} obtains $f = 0.32$
\cite{Giedt:2009mr}.  These do not reflect the full range of
possible values , which ref.\ \cite{Mambrini:2011ik} puts at 
$f=0.26-0.63$, corresponding to a factor of 6 uncertainty in the
direct detection cross section.  We
adopt the median value $f=0.35$ for definiteness, but one might 
reasonably invoke $f=0.26$ to weaken the effect of
direct detection limits (by a factor of $1.8$ in the cross section) 
and thus the degree of fine-tuning of 
model parameters that we will find below.

Because the same Feynman diagram
is responsible for both processes (\ref{sigma_relic}) and (\ref{sigma_dd}), 
the XENON100 constraint restricts $\lambda_{DM}$ to be small.  Thus to get a large
enough annihilation cross section  (\ref{sigma_relic}), we must be somewhat close to
the resonance condition $m_{DM}\cong m_h/2$, and the model thus requires some 
moderate tuning.  We will quantify this below.  The XENON100 90\% c.l.\ limit is 
$\sigma_{SI} \lesssim 8\times 10^{-45}$ cm$^2$ in the DM mass range of interest, assuming that
$m_h \cong 126$ GeV.

One further condition  we impose is that invisible decays
of the SM Higgs boson $h\to HH$ do not dominate its width, where
the invisible contribution is given by
\be
\Gamma_{\rm inv} = {\lambda^2_{DM} v^2\over 64 \pi m_h} 
\sqrt{1-4\,m^2_{DM}/m^2_h}
\ee
with SM Higgs VEV $v=246$ GeV.  Ref.\
\cite{Bai:2011wz} indicates that a constraint at the level of 40\% 
on the branching ratio for such decays should be
attainable in the relatively near future 
from LHC data.   Although not yet established, it seems unlikely that
the invisible decays dominate the width of the Higgs if the excess
events seen at LHC are really due to the Higgs, so we provisionally impose the
40\% constraint.   Below we will show that this requirement restricts
the range of allowed $m_{DM}$, but only slightly more than the 
combination of relic density and direct detection constraints.

We also explore a generalization of the above scenario, in which the
$H$ boson could make a subdominant contribution $\Omega_{H}$ to the total dark
matter density.  This will occur if $\lambda_{DM}$ is larger, for a
given value of $m_{DM}$, than what is required to satisfy
(\ref{wmap}).  The constraint on $\sigma_{SI}$ from direct detection
is correspondingly weakened since the rate goes like 
$\Omega_{H}\,\sigma_{SI}\sim \sigma_{SI}/\langle\sigma v\rangle$.
The factors of $\lambda_{DM}$ cancel out and thus the combined
constraints become independent of $\lambda_{DM}$ so long as it is
large enough to sufficiently suppress the relic density.

\section{Monte Carlo Results}
\label{results}

\begin{figure}[tb]
\centering
\includegraphics[width=0.5\textwidth]{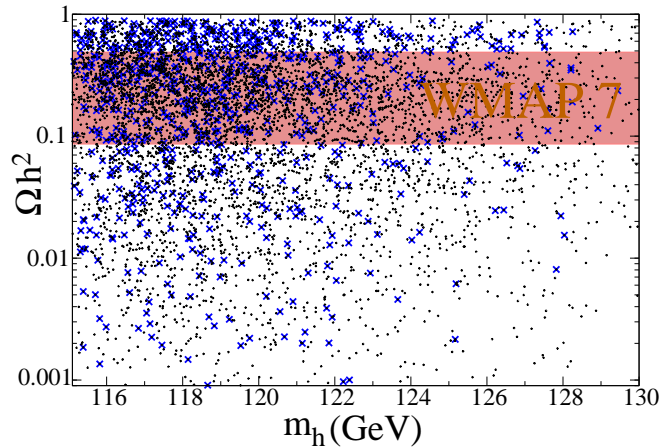}
\caption{Scatter plot of dark matter relic density $\Omega h^2$ versus
Higgs mass $m_h$ from Monte Carlo, for models with a 
strong first order electroweak phase transition 
($v_c/T_c>1$). Dense points (black) correspond to models that
do not necessarily satisfy the
XENON100 constraint, while crosses (blue) indicate models that do.
The shaded band corresponds to relic density 
$\Omega h^2 \in [0.085, 0.139] $ observed by WMAP at $3\sigma$.}
\label{fig1}
\end{figure}

We initially performed a MCMC scan of the full model parameter space 
(varying $\lambda_1$, $\lambda_2$, $\lambda_3$, $\lambda_S$, $m_1^2$)
as described above, without imposing any constraints upon the relic
density or direct detection cross section. We find in this way many
models that satisfy the sphaleron constraint $v_c/T_c > 1$ on the
ratio of the Higgs VEV to the critical temperature.  These results are
illustrated in fig.\ \ref{fig1} which plots  the DM relic density
versus SM-like Higgs mass $m_h$ for the models with a strong phase
transition (black points).   It can be seen that there are many such
examples within the  mass window $m_h = 115-130 \; \text{GeV}$, as
well as within the $3\sigma$ allowed range $\Omega h^2 \in [0.085,
0.139]$ for the relic density as determined by  Seven-Year Wilkinson
Microwave Anisotropy Probe (WMAP) observations \cite{Jarosik:2010iu},
and indicated by the shaded horizontal band.   A fraction of the
points also satisfy the XENON100 constraint; these are denoted by blue
crosses. A small population can be found that
simultaneously satisfy both constraints and which have $m_h$ near 126
GeV.  We use these as seeds for a more focused MCMC search in
which only models with the correct relic density and small enough
direct detection cross section are admitted into the chains.

\begin{figure}[tb]
\centering
\includegraphics[width=0.48\textwidth]{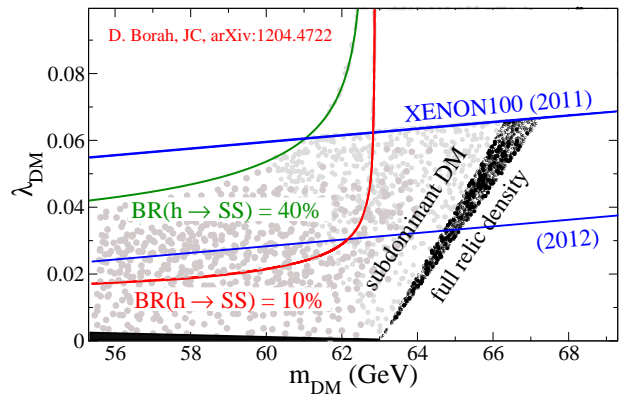}
\caption{Scatter plot of  $\lambda_{DM}$ versus $m_{DM}$ for models
with strong EWPT, correct relic density (dark points), and $m_h=126$ GeV. The 90\%
c.l.\ upper bounds on 
$\lambda_{DM}$ from XENON100 (2011) \cite{Aprile:2011hi} and (2012) 
\cite{Aprile:2012nq} are shown by the 
slanted lines.  The
light shaded points denote models whose relic density is subdominant,
$\Omega_H h^2 < 0.085$, but which still satisfy the correspondingly
relaxed XENON100 limit.  
The other curves indicate the upper limit 
on $\lambda_{DM}$ from requiring that the branching ratio
for the invisible decay
$h \rightarrow HH $  not exceed 10\% or 40\%,
respectively.}
\label{fig2}
\end{figure}

\begin{figure}[tb]
\centering
\includegraphics[width=0.5\textwidth]{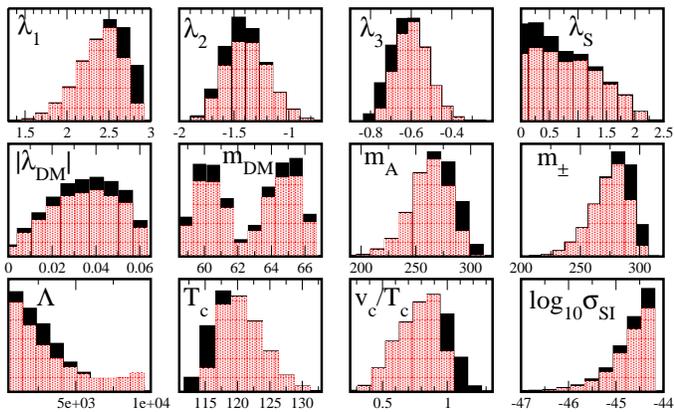}
\caption{Monte Carlo distributions of model parameters and derived quantities satisfying
all constraints (including $\Omega_{H}h^2\in [0.085,0.139]$), with $m_h$ fixed at 126 GeV.  Light-shaded regions
indicate the proportion of frequency contributed by models with
$v_c/T_c < 1$, while dark corresponds to $v_c/T_c > 1$.
$\Lambda$ is
the energy scale of the Landau pole for each model.  
Masses, $\Lambda$  and the
critical temperature $T_c$ are in GeV units.  The spin-independent DM-nucleon
scattering cross section $\sigma_{SI}$ is in units of cm$^2$.
 }
\label{fig3}
\end{figure}

Highlighting the effect of the direct detection constraint, figure
\ref{fig2} shows the scatter plot of MCMC models having $v_c/T_c>1$
and correct relic density,\footnote{We have corrected fig.\
\ref{fig2} since the original submission by approximately taking into
account the effect of thermal averaging, not done in eq.\ 
(\ref{sigma_relic}),
which is the cross section in the limit of zero velocity.  Thermal
averaging allows for some fraction of DM with $m_{DM}<m_h/2$ to
cause resonant annihilation and so significantly suppresses the relic
density relative to using eq.\ (\ref{sigma_relic}) in these cases.} 
 in the  plane of $\lambda_{DM}$ versus
$m_{DM}$ for the case of $m_h = 126 \; \text{GeV}$ which is suggested
by the recent results from  the ATLAS and CMS experiments.  The
XENON100 upper limit on $\lambda_{DM}$ is plotted as the slanting
line.  The peculiar V-shape of the allowed region is due to the need
for being close to resonance of the virtual Higgs boson in the
$s$-channel to get sufficiently strong annihilation for the relic
abundance.  Larger values of
$\lambda_{DM}$ do not require $m_{DM}$ to be as close to $m_h/2$. 
The different density of points above and below the Xenon constraint
is due to using a rough, preliminary MCMC to find the former points,
and the more focused search to find the latter.  

We also show in fig.\ \ref{fig2} the
upper limits on $\lambda_{DM}$ from requiring that the branching
ratio for the invisible decays BR($h\to HH)$ not exceed 40\% or 10\%
respectively.  These are futuristic requirements, but
ref.\ \cite{Bai:2011wz} estimates that the
40\% limit will be attainable with 20 fb$^{-1}$ of integrated 
luminosity at the LHC. When combined with the allowed region
in our plot this constraint leads to the lower bound $m_{DM}\gtrsim 60$ GeV.  This
is only slightly more stringent than the bound due to direct
detection on the left arm of the ``V''.  The interior region of 
the ``V'' is populated by models for which $H$ makes a subdominant
contribution to the relic density.  In this sample, the constraint
on the invisible width of the Higgs can be more important than 
the XENON100 constraint, over a wider range of $H$ masses.

A summary of the favored values of the Lagrangian parameters, mass
spectra, and EWPT and DM attributes, from the refined MCMC search
implementing all constraints (assuming $H$ accounts for all of
the DM and not just a subdominant component), 
 is given by the histogram plots of
fig.\ \ref{fig3}, where the light-shaded portions of the
bars indicate the fraction of models in the chain having 
$v_c/T_c<1$ for the given parameter value, while the dark shaded
part shows the proportion contributed by models with $v_c/T_c>1$.
The chain contains 14400 models, of which 2400 have $v_c/T_c>1$,
despite the biasing of the MCMC toward large values of $v_c/T_c$.
This shows that obtaining a strong first order EWPT is not altogether
easy.

Another striking feature is that $\lambda_1$ is always
large for the cases with $v_c/T_c > 1$, 
in the range $2.6-3$.  One might expect such large couplings to
cause a breakdown of perturbation theory.  Nevertheless we have run
the renormalization group equations up to find the Landau pole energy
scale $\Lambda$ in each case, keeping only those models with $\Lambda
> 2$ TeV. The values of $\Lambda$ for our accepted models are in the
range $2-5$ TeV; thus some additional new physics must come into play
at these higher energies.  

A consequence of the large values of $\lambda_1$ and $|\lambda_2|$ is
that there must be fine-tuning between $\lambda_1$ and the combination
$\lambda_2 + 2\lambda_3$ at the level of $(1.4\pm 0.5)$\% (the minimum level of
tuning we find is 2.5\% among the models in our chain with $v_c/T_c >
1$) in order to make $\lambda_{DM}$ sufficiently small.  As we noted
above, there is an additional tuning between $2 m_{DM}$ and $m_h$,
which we find to be at the level of $(3.7\pm 1.4)$\% in the $v_c/T_c >
1$ subsample.  The fine-tuning problem for $\lambda_{DM}$ is slightly
ameliorated in the scan of models having $\Omega_{H}h^2 < 0.085$, for
which we find that $\lambda_2+2\lambda_3$ must cancel $\lambda_1$
at the level of $3.8\pm 2.7\%$, a factor of 3 improvement.

\begin{figure}[tb]
\centering
\includegraphics[width=0.5\textwidth]{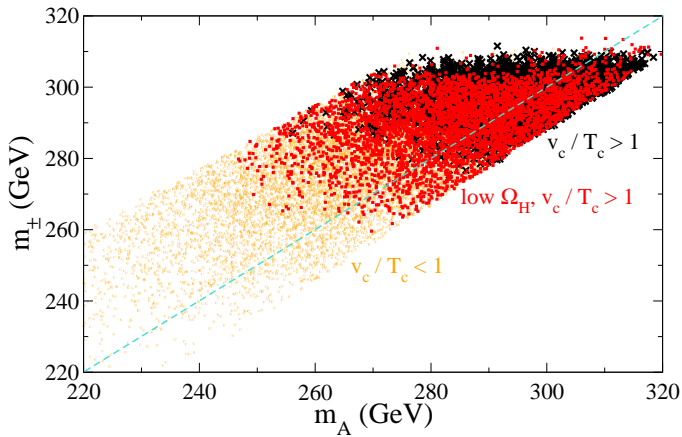}
\caption{Scatter plot of charged Higss mass $m_\pm$ versus CP-odd 
Higgs mass $m_A$, for models satisfying all constraints.   
Dark-shaded upper region (crosses): models with
with $v_c/T_c >1$; lighter-shaded middle region (squares): models with
subdominant DM component $\Omega_H h^2 < 0.085$ but $v_c/T_c >1$;
lightest-shaded lower region (dots): models with $v_c/T_c <1$.
The diagonal line shows where $m_\pm = m_A$.
 }
\label{fig4}
\end{figure}

\section{Implications for LHC}
\label{lhc}

The IDM presents a challenge for discovery at
collider experiments.   
From fig.\ \ref{fig3}, we see that  the masses of the CP-odd and
charged Higgs bosons $A$ and $H^\pm$ are predicted to be in  a relatively narrow
window $260-320$ GeV.  We find that there is no strong correlation
between $m_A$ or $m_\pm$ and  $m_{DM}$, but there is a
 noticeable correlation (due
to the EWPO constraint) between  $m_\pm$ and $m_A$, shown in fig.\
\ref{fig4}.   The allowed region consistent with $v_c/T_c>1$ 
and $H$ providing the dominant DM component is
shown by the dark crosses in the upper part of the region of
interest.  This gets extended to somewhat smaller masses for the
case of models where $H$ is a subdominant DM component.

Because the inert doublet does not couple to quarks, its
production cross section is suppressed relative to that of the SM
Higgs, proceeding mainly by $q\bar q\to  (A,H)$, $(H^\pm,H^\mp)$,
$(H^\pm,A)$ or $(H^\pm,H)$ through a virtual $Z$ or $W$ in the $s$ channel.
Ref.\ \cite{Dolle:2009ft} argues that a promising discovery
signal is through the subsequent decay of the heavy Higgs boson(s)
to produce lepton pairs and missing energy.  However the models shown
to be most favorable for discovery with 100 fb$^{-1}$ of integrated
luminosity are those with $|m_A-m_{DM}| \sim 40-80$ GeV or 
$m_{DM}\sim 40$ GeV, which do not include the ones we predict in
the present analysis.  A similar conclusion holds for the trilepton
discovery channel \cite{Miao:2010rg}.

Ref.\ \cite{Fox:2011pm} more generally considers monojet events
induced by dimension 6 operators of the form $\bar
q q\bar\chi\chi/\Lambda^2$ for Dirac DM $\chi$ coupling to quarks,
producing a jet from initial state radiation of a gluon from
a quark, as well as missing energy from the DM.
The IDM generates such operators with a virtual $Z$ boson
at one loop, through the $ZZHH$ coupling, leading to a dimension 6
operator of order  
$(G_F/96\pi^2)\bar q\!\!\lrasd\!\! q |H|^2$.
Even though the effective 
description is not valid at LHC energies,
it should give an upper bound on the strength of this virtual process. 
Due to the initial state radiation,
one of the quarks is off-shell, so $\slashed\partial$ will be of order
the momentum of the jet rather than the quark mass.  Even so, we find
that the $\bar q\!\!\lrasd\!\! q |H|^2$
operator  produces an amplitude similar to that of 
$\bar q q\bar\chi\chi/\Lambda^2$ with $\Lambda\sim 7$ TeV, which
is far beyond the current sensitivity of LHC experiments, $\Lambda > 
700$ GeV, determined by ref.\ \cite{Fox:2011pm}.  The mismatch
arises because of the loop suppression factor in our model.  The
monojet constraint is more stringent for operators generated by
tree-level exchange.  For the IDM, the most important such process
is $q\bar q\to HA$ via $s$-channel $Z$ exchange, followed by 
$A\to H Z$.    Hadronic
decays $Z\to q\bar q H$ could then appear as monojets plus missing
energy, if the $Z$ is sufficiently boosted so that only one jet is
resolved.  It could be interesting to undertake a study of such
processes tailored to the IDM.

Another possibility for testing the IDM is by the effect of the 
charged Higgses on the $h\to\gamma\gamma$ branching ratio
\cite{Posch:2010hx,Arhrib:2012ia}.  The rate for $h\to\gamma\gamma$
is given by
\be
	\Gamma = {G_F \alpha^2 m_h^3\over 128\sqrt{2}\pi^3}\,|A_{SM} +
	A_{H^\pm}|^2
\ee
where $A_{SM} = -6.52$ for $m_h=126$ GeV and 
\be
	A_{H^+} = -{\lambda_1 v^2\over 2 m_\pm^2}
\left(\tau^{-1}-\tau^{-2}(\sin^{-1}\sqrt{\tau})^2\right) \cong \frac13
\ee	
with $\tau= (m_h/2m_\pm)^2$.  The approximation $A_{H^+} \cong 1/3$
holds in the limit $m_\pm \gg m_h$ and $\lambda_1 v^2 \cong 2 m^2_\pm$
which are satisfied in the models favored by our analysis.  Therefore
the $H^\pm$ contribution interferes destructively with that of the SM,
and results in a increase close to $10\%$ in the $h\to\gamma\gamma$ 
partial width for all the models that we consider.  However there is
a larger effect on the branching ratio for $h\to\gamma\gamma$ due to
the invisible decay channel $h\to HH$, which increases the total decay
width relative to that in the standard model.  Thus the change 
in the branching ratio BR($h\to\gamma\gamma$) may be
dominated by the dilution due to $h\to HH$ in the region where
$m_H < m_h/2$.  If on the other hand $m_H > m_h/2$, we have the
definite prediction that BR($h\to\gamma\gamma$) is close to 90\%
of its standard model value.  Such a reduction
is in the opposite direction of the upward fluctuation in the value
that was previously observed by CMS \cite{Chatrchyan:2012tx}.

\section{Conclusions}

We have quantitatively reconsidered the impact of an inert Higgs
doublet on the strength of the electroweak phase transition, generally
confirming the result of ref.\ \cite{Chowdhury:2011ga} that it is
relatively easy to find models satisfying all constraints and giving
$v_c/T_c>1$, but we differ in the details.  Whereas ref.\ 
\cite{Chowdhury:2011ga} finds allowed masses  $m_\pm \cong m_A$ in the
range $280-370$ GeV, we find a more restricted range $260-320$ GeV,
correlated with our limit $\lambda_1 < 3$ in contrast to theirs,
$\lambda_1 < 4$.   Considering the approximate nature of the
finite-temperature effective potential used in
\cite{Chowdhury:2011ga}, the results seem to be in reasonable
agreement.  

However, we have pointed out some fine-tunings needed to make the
scenario work: $m_{DM}$ must be within $\sim 4$ GeV of $m_h/2$ to get
a strong enough  $HH\to b\bar b$ annihilation cross section for the
observed relic density, while the large value of $\lambda_1$ needed to
get $v_c/T_c >1$ is canceled typically at the (1-4)\% level 
(depending upon whether $H$ is the dominant component of the 
total DM density) by $\lambda_2
+ 2\lambda_3$ so that the DM-nucleon coupling is small enough to
satisfy the XENON100 constraint.  If the IDM dark matter plus
strong EWPT scenario
should be borne out by experiments, it will be mysterious why  these
two seemingly unlikely coincidences should exist.

We have considered the prospects for testing the scenario at the LHC. 
There are several possible signatures: invisible decays of the SM-like
Higgs into dark matter could be inferred if $m_{DM} \lesssim 61$ GeV. 
A 10\% decrease in the partial width for $h\to\gamma\gamma$ is a firm
prediction.  Current analyses of missing energy plus monojets or
dileptons do not seem able to rule out the model in the near future,
but we suggest for further study that the process $q\bar q\to H A$
followed by $A\to Z H$ and hadronic decays of the  $Z$ (if
sufficiently boosted) could give monojet-like events that might
constitute a more promising signal.

\label{conclusions}

{\bf Acknowledgements}.  We thank Mike Trott for his kind assistance in generating
EWPO constraints for the mass ranges of interest and for valuable
suggestions, and Joel Giedt, Joachim Kopp, Sabine Kraml and Guy Moore for 
enlightening discussions.
The visit of D.B.\ to McGill
University was supported by Canadian Commonwealth Fellowship Program.
JC's research is supported by NSERC (Canada).

\bibliographystyle{apsrev}

\begin{thebibliography}{10}
\expandafter\ifx\csname natexlab\endcsname\relax\def\natexlab#1{#1}\fi
\expandafter\ifx\csname bibnamefont\endcsname\relax
  \def\bibnamefont#1{#1}\fi
\expandafter\ifx\csname bibfnamefont\endcsname\relax
  \def\bibfnamefont#1{#1}\fi
\expandafter\ifx\csname citenamefont\endcsname\relax
  \def\citenamefont#1{#1}\fi
\expandafter\ifx\csname url\endcsname\relax
  \def\url#1{\texttt{#1}}\fi
\expandafter\ifx\csname urlprefix\endcsname\relax\def\urlprefix{URL }\fi
\providecommand{\bibinfo}[2]{#2}
\providecommand{\eprint}[2][]{\url{#2}}

\bibitem{Anderson:1991zb} 
  G.~W.~Anderson and L.~J.~Hall,
  Phys.\ Rev.\ D {\bf 45}, 2685 (1992).


\bibitem{Dimopoulos:1990ai} 
  S.~Dimopoulos, R.~Esmailzadeh, L.~J.~Hall and N.~Tetradis,
  Phys.\ Lett.\ B {\bf 247}, 601 (1990).


\bibitem{Barger:2008jx} 
  V.~Barger, P.~Langacker, M.~McCaskey, M.~Ramsey-Musolf and G.~Shaughnessy,
  Phys.\ Rev.\ D {\bf 79}, 015018 (2009)
  [arXiv:0811.0393 [hep-ph]].

\bibitem{Kang:2009rd} 
  J.~Kang, P.~Langacker, T.~Li and T.~Liu,
  JHEP {\bf 1104}, 097 (2011)
  [arXiv:0911.2939 [hep-ph]].

\bibitem{Kumar:2011np} 
  P.~Kumar and E.~Ponton,
  JHEP {\bf 1111}, 037 (2011)
  [arXiv:1107.1719 [hep-ph]].

\bibitem{Chung:2011it} 
  D.~J.~H.~Chung and A.~J.~Long,
  Phys.\ Rev.\ D {\bf 84}, 103513 (2011)
  [arXiv:1108.5193 [astro-ph.CO]].

\bibitem{Kozaczuk:2011vr} 
  J.~Kozaczuk and S.~Profumo,
  JCAP {\bf 1111}, 031 (2011)
  [arXiv:1108.0393 [hep-ph]].

\bibitem{Carena:2011jy} 
  M.~Carena, N.~R.~Shah and C.~E.~M.~Wagner,
  Phys.\ Rev.\ D {\bf 85}, 036003 (2012)
  [arXiv:1110.4378 [hep-ph]].

\bibitem{Ahriche:2012ei} 
  A.~Ahriche and S.~Nasri,
  arXiv:1201.4614 [hep-ph].


\bibitem{Gonderinger:2012rd} 
  M.~Gonderinger, H.~Lim and M.~J.~Ramsey-Musolf,
  arXiv:1202.1316 [hep-ph].

\bibitem{Ma:2006km} 
  E.~Ma,
  Phys.\ Rev.\ D {\bf 73}, 077301 (2006)
  [hep-ph/0601225].


\bibitem{Barbieri:2006dq} 
  R.~Barbieri, L.~J.~Hall and V.~S.~Rychkov,
  Phys.\ Rev.\ D {\bf 74}, 015007 (2006)
  [hep-ph/0603188].

\bibitem{Majumdar:2006nt} 
  D.~Majumdar and A.~Ghosal,
  Mod.\ Phys.\ Lett.\ A {\bf 23}, 2011 (2008)
  [hep-ph/0607067].

\bibitem{LopezHonorez:2006gr} 
  L.~Lopez Honorez, E.~Nezri, J.~F.~Oliver and M.~H.~G.~Tytgat,
  JCAP {\bf 0702}, 028 (2007)
  [hep-ph/ 0612275].

\bibitem{Cao:2007rm} 
  Q.~-H.~Cao, E.~Ma and G.~Rajasekaran,
  Phys.\ Rev.\ D {\bf 76}, 095011 (2007)
  [arXiv:0708.2939 [hep-ph]].

\bibitem{Agrawal:2008xz} 
  P.~Agrawal, E.~M.~Dolle and C.~A.~Krenke,
  Phys.\ Rev.\ D {\bf 79}, 015015 (2009)
  [arXiv:0811.1798 [hep-ph]].


\bibitem{Andreas:2009hj} 
  S.~Andreas, M.~H.~G.~Tytgat and Q.~Swillens,
  JCAP {\bf 0904}, 004 (2009)
  [arXiv:0901.1750 [hep-ph]].

\bibitem{Nezri:2009jd} 
  E.~Nezri, M.~H.~G.~Tytgat and G.~Vertongen,
  JCAP {\bf 0904}, 014 (2009)
  [arXiv:0901.2556 [hep-ph]].

\bibitem{Dolle:2009ft} 
  E.~Dolle, X.~Miao, S.~Su and B.~Thomas,
  Phys.\ Rev.\ D {\bf 81}, 035003 (2010)
  [arXiv:0909.3094 [hep-ph]].


\bibitem{Arina:2009um} 
  C.~Arina, F.~-S.~Ling and M.~H.~G.~Tytgat,
  JCAP {\bf 0910}, 018 (2009)
  [arXiv:0907.0430 [hep-ph]].

\bibitem{Posch:2010hx} 
  P.~Posch,
  Phys.\ Lett.\ B {\bf 696}, 447 (2011)
  [arXiv:1001.1759 [hep-ph]].

\bibitem{Honorez:2010re} 
  L.~Lopez Honorez and C.~E.~Yaguna,
  JHEP {\bf 1009}, 046 (2010)
  [arXiv:1003.3125 [hep-ph]].

\bibitem{Miao:2010rg} 
  X.~Miao, S.~Su and B.~Thomas,
  Phys.\ Rev.\ D {\bf 82}, 035009 (2010)
  [arXiv:1005.0090 [hep-ph]].

\bibitem{Melfo:2011ie} 
  A.~Melfo, M.~Nemevsek, F.~Nesti, G.~Senjanovic and Y.~Zhang,
  Phys.\ Rev.\ D {\bf 84}, 034009 (2011)
  [arXiv:1105.4611 [hep-ph]].

\bibitem{Arina:2011cu} 
  C.~Arina and N.~Sahu,
  Nucl.\ Phys.\ B {\bf 854}, 666 (2012)
  [arXiv:1108.3967 [hep-ph]].

\bibitem{Arhrib:2012ia} 
  A.~Arhrib, R.~Benbrik and N.~Gaur,
  arXiv:1201.2644 [hep-ph].


\bibitem{Chowdhury:2011ga} 
  T.~A.~Chowdhury, M.~Nemevsek, G.~Senjanovic and Y.~Zhang,
  JCAP {\bf 1202}, 029 (2012)
  [arXiv:1110.5334 [hep-ph]].

\bibitem{Cline:2011mm} 
  J.~M.~Cline, K.~Kainulainen and M.~Trott,
  JHEP {\bf 1111}, 089 (2011)
  [arXiv:1107.3559 [hep-ph]].

\bibitem{ATLAS:2012ae} 
  G.~Aad {\it et al.}  [ATLAS Collaboration],
  Phys.\ Lett.\ B {\bf 710}, 49 (2012)
  [arXiv:1202.1408 [hep-ex]].

\bibitem{Chatrchyan:2012tx} 
  S.~Chatrchyan {\it et al.}  [CMS Collaboration],
  arXiv: 1202.1488 [hep-ex].

\bibitem{Chatrchyan:2012tw} 
  S.~Chatrchyan {\it et al.}  [CMS Collaboration],
  arXiv: 1202.1487 [hep-ex].

\bibitem{ATLAS:2012ad} 
  G.~Aad {\it et al.}  [ATLAS Collaboration],
  Phys.\ Rev.\ Lett.\  {\bf 108}, 111803 (2012)
  [arXiv:1202.1414 [hep-ex]].


\bibitem[{\citenamefont{Cirelli et~al.}(2006)\citenamefont{Cirelli, Fornengo,
  and Strumia}}]{Cirelli:2005uq}
\bibinfo{author}{\bibfnamefont{M.}~\bibnamefont{Cirelli}},
  \bibinfo{author}{\bibfnamefont{N.}~\bibnamefont{Fornengo}}, \bibnamefont{and}
  \bibinfo{author}{\bibfnamefont{A.}~\bibnamefont{Strumia}},
  \bibinfo{journal}{Nucl. Phys.} \textbf{\bibinfo{volume}{B753}},
  \bibinfo{pages}{178} (\bibinfo{year}{2006}), \eprint{hep-ph/0512090}.


\bibitem[{\citenamefont{Lundstrom et~al.}(2009)\citenamefont{Lundstrom,
  Gustafsson, and Edsjo}}]{Lundstrom:2008ai}
\bibinfo{author}{\bibfnamefont{E.}~\bibnamefont{Lundstrom}},
  \bibinfo{author}{\bibfnamefont{M.}~\bibnamefont{Gustafsson}},
  \bibnamefont{and} \bibinfo{author}{\bibfnamefont{J.}~\bibnamefont{Edsjo}},
  \bibinfo{journal}{Phys. Rev.} \textbf{\bibinfo{volume}{D79}},
  \bibinfo{pages}{035013} (\bibinfo{year}{2009}), \eprint{0810.3924}.



\bibitem{Pierce:2007ut} 
  A.~Pierce and J.~Thaler,
  JHEP {\bf 0708}, 026 (2007)
  [hep-ph/0703056 [HEP-PH]].


\bibitem{Cline:2006ts} 
  J.~M.~Cline,
  ``Baryogenesis,'' lectures given at Les Houches Summer School
  2006, 
  hep-ph/0609145.

\bibitem[{\citenamefont{Jungman et~al.}(1996)\citenamefont{Jungman,
  Kamionkowski, and Griest}}]{Jungman:1995df}
\bibinfo{author}{\bibfnamefont{G.}~\bibnamefont{Jungman}},
  \bibinfo{author}{\bibfnamefont{M.}~\bibnamefont{Kamionkowski}},
  \bibnamefont{and} \bibinfo{author}{\bibfnamefont{K.}~\bibnamefont{Griest}},
  \bibinfo{journal}{Phys. Rept.} \textbf{\bibinfo{volume}{267}},
  \bibinfo{pages}{195} (\bibinfo{year}{1996}), \eprint{hep-ph/9506380}.



\bibitem[{\citenamefont{Jarosik et~al.}(2011)}]{Jarosik:2010iu}
\bibinfo{author}{\bibfnamefont{N.}~\bibnamefont{Jarosik}} \bibnamefont{et~al.},
  \bibinfo{journal}{Astrophys. J. Suppl.} \textbf{\bibinfo{volume}{192}},
  \bibinfo{pages}{14} (\bibinfo{year}{2011}), \eprint{1001.4744}.

\bibitem[{\citenamefont{Aprile et~al.}(2010)}]{Aprile:2010um}
\bibinfo{author}{\bibfnamefont{E.}~\bibnamefont{Aprile}} \bibnamefont{et~al.}
  (\bibinfo{collaboration}{XENON100}), \bibinfo{journal}{Phys. Rev. Lett.}
  \textbf{\bibinfo{volume}{105}}, \bibinfo{pages}{131302}
  (\bibinfo{year}{2010}), \eprint{1005.0380}.


\bibitem{Aprile:2011hi} 
  E.~Aprile {\it et al.}  [XENON100 Collaboration],
  Phys.\ Rev.\ Lett.\  {\bf 107}, 131302 (2011)
  [arXiv:1104.2549 [astro-ph.CO]].

\bibitem{Aprile:2012nq} 
  E.~Aprile {\it et al.}  [XENON100 Collaboration],
  Phys.\ Rev.\ Lett.\  {\bf 109}, 181301 (2012)
  [arXiv:1207.5988 [astro-ph.CO]].


\bibitem{Ellis:2008hf} 
  J.~R.~Ellis, K.~A.~Olive and C.~Savage,
  Phys.\ Rev.\ D {\bf 77}, 065026 (2008)
  [arXiv:0801.3656 [hep-ph]].

\bibitem{Bottino:2008mf} 
  A.~Bottino, F.~Donato, N.~Fornengo and S.~Scopel,
  Phys.\ Rev.\ D {\bf 78}, 083520 (2008)
  [arXiv:0806.4099 [hep-ph]].

\bibitem{Mambrini:2011ik} 
  Y.~Mambrini,
  Phys.\ Rev.\ D {\bf 84}, 115017 (2011)
  [arXiv:1108.0671 [hep-ph]].

\bibitem{Gondolo:2004sc} 
  P.~Gondolo, J.~Edsjo, P.~Ullio, L.~Bergstrom, M.~Schelke and E.~A.~Baltz,
  JCAP {\bf 0407}, 008 (2004)
  [astro-ph/0406204].

\bibitem{Ellis:2000ds} 
  J.~R.~Ellis, A.~Ferstl and K.~A.~Olive,
  Phys.\ Lett.\ B {\bf 481}, 304 (2000)
  [hep-ph/0001005].


\bibitem[{\citenamefont{Giedt et~al.}(2009)\citenamefont{Giedt, Thomas, and
  Young}}]{Giedt:2009mr}
\bibinfo{author}{\bibfnamefont{J.}~\bibnamefont{Giedt}},
  \bibinfo{author}{\bibfnamefont{A.~W.} \bibnamefont{Thomas}},
  \bibnamefont{and} \bibinfo{author}{\bibfnamefont{R.~D.} \bibnamefont{Young}},
  \bibinfo{journal}{Phys. Rev. Lett.} \textbf{\bibinfo{volume}{103}},
  \bibinfo{pages}{201802} (\bibinfo{year}{2009}), \eprint{0907.4177}.

\bibitem{Bai:2011wz} 
  Y.~Bai, P.~Draper and J.~Shelton,
  arXiv:1112.4496 [hep-ph].

\bibitem{Fox:2011pm} 
  P.~J.~Fox, R.~Harnik, J.~Kopp and Y.~Tsai,
  arXiv: 1109.4398 [hep-ph].


\end{thebibliography}

\end{document}